\documentclass[sn-mathphys]{sn-jnl}

\usepackage{caption}
\jyear{2024}%

\theoremstyle{thmstyleone}%
\theoremstyle{thmstyletwo}%

\theoremstyle{thmstylethree}%

\begin{document}

\title[]{A metamaterial telescope at millimetre wavelengths.}

\author*[1]{\fnm{Giorgio} \sur{Savini}}\email{g.savini@ucl.ac.uk}

\author[2]{\fnm{Peter} \sur{Hargrave}}\email{HargravePC@cardiff.ac.uk}

\author[2]{\fnm{Peter A.R.} \sur{Ade}}\email{peter.ade@astro.cf.ac.uk}

\author[1]{\fnm{Alexey} \sur{Shitvov}}\email{a.shitvov@ucl.ac.uk}

\author[2]{\fnm{Rashmi} \sur{Sudiwala}}\email{r.sudiwala@astro.cf.ac.uk}

\author[3]{\fnm{Giampaolo} \sur{Pisano}}\email{giampaolo.pisano@uniroma1.it}

\author[2]{\fnm{Carole} \sur{Tucker}}\email{carole.tucker@astro.cf.ac.uk}

\author[4]{\fnm{Jin} \sur{Zhang}}\email{jin.zhang@aru.ac.uk}

\affil*[1]{\orgdiv{Physics \& Astronomy Dept.}, \orgname{University College London}, \orgaddress{\street{Gower Street}, \city{London}, \postcode{WC1E 6BT}, \country{UK}}}

\affil[2]{\orgdiv{School of Physics \& Astronomy}, \orgname{Cardiff University}, \orgaddress{\street{The Parade}, \city{Cardiff}, \postcode{CB24 3AA}, \state{Wales}, \country{UK}}}

\affil[3]{\orgdiv{Dip. Fisica}, \orgname{Univ. Roma "La Sapienza"}, \orgaddress{\street{P.le Aldo Moro}, \city{Roma}, \postcode{00100}, \country{Italy}}}

\affil[4]{\orgdiv{School of Computing and Information Science}, \orgname{Anglia Ruskin University}, \orgaddress{\street{East Road}, \city{Cambridge}, \postcode{CB1 1PT}, \country{UK}}}


\abstract{In this paper we present a novel telescope composed exclusively of thin, flat optical elements, each being a hot-pressed multi-layered structure combining the properties of a lens, its anti-reflection coating and frequency selection or filtering. We discuss the design process, from fundamental physical metamaterial properties of the single periodic cell structure to the lens concept, which constitutes the building block of the telescope design, and the iterative process that is part of the lens optimization. We provide the results of a laboratory test campaign for different telescope designs based on three-lens arrangements. Beam cuts and focus measurements both on- and off-axis are compared with models showing good agreement. We conclude that a broad-band mm-wave complete telescope system consisting entirely of metamaterial flat lenses has been built and tested, showing comparable performance with conventional state-of-the-art refractive telescopes in the same wavelength region. 
This new broadband design, highly efficient at frequencies between 90 and 190 GHz, offers multiple advantages.  These include a $> 80\%$ weight reduction, reduced issues tied to coating-survivability at cryogenic temperatures caused by differential contraction exacerbated by non-flat surfaces, as well as a reduction in the overall number of components and mechanical mounts.}



\maketitle

\section{Introduction}\label{sec1}

Wide-field telescopes are used extensively in far-infrared and mm-wavelength astronomy. 
Covering a wavelength range between 1 mm and 1 cm or an equivalent frequency range between 30 and 300 GHz, such telescopes have been used, both in astrophysics and cosmology experiments \cite{Planck1}, \cite{Simons2024}, \cite{Ghigna2024} as well as Earth Observing \cite{Yang2011}, \cite{Ilias2013}.

In the former, the spectral range considered includes the peak of the cosmic microwave background (CMB) as well as a number of its foreground signal contaminants. In Earth Observing, the same frequency range is used to monitor oxygen and water lines useful for weather forecasting and climate science. The lower part of that same spectral range has also become an experimentation ground for telecommunications both on the ground and in space \cite{Jornet2023}, \cite{ofcom2020}.

The need for wide field coverage both on sky, for astrophysics studies, and in Earth observing, for wide swath coverage, in turn dictates fast optics and thick refractive lenses as an alternative to off-axis reflective optics. 
Conventional refractive telescope lens systems adopted in current and planned satellite-based microwave instruments are unavoidably large in area, 
but also thick quasi-optical components\cite{Ghigna2024}.

In addition to higher absorption losses, thick and high-curvature lenses present a range of manufacturing issues when one or more anti-reflection coating (ARC) layers are added to minimize reflection loss. These issues are exasperated when operating such telescopes in a cryogenic vacuum or Space, due to coating delamination or thermo-elastic distortions.
A second requirement for such telescopes is to serve a wide range of frequencies be that by hosting multi-chroic detector focal planes \cite{Ghigna2024} or by combining multiple receivers in complex dichroic-split optical path instruments \cite{Lupi2020}. This has led in some cases to prefer reflective telescopes even though the mass and volume required in this case can be substantially larger (with added complexities). Bandwidth limitations and manufacturing constraints for refractive components mean that reflective optical solutions are the traditional choice for large aperture telescopes.

In 2008 we envisioned making use of a tried and tested technology for manufacturing frequency selective filters in this same frequency range \cite{Ade2006} by considering a metamaterial composed by the same kind of photolithographic grid-stacks and combining them with the old optical design of flat Gradient Index (or "Wood") lenses\cite{Wood1905}  which these days are quite common in the visible wavelength domain especially in fibre optics.
The relative ease in using this technology for tuning the index of refraction was initially first proven in \cite{Zhang2009} by tuning the index along the optical axis to be used as ARC and adopted successfully in cryo-vacuum conditions on different experiments in the field (\cite{Pascale2012} and \cite{Bernard2016}]). New devices using the same technology were subsequently achieved with a non-periodic distribution on a plane orthogonal to the optical axis for use for a single lens designs (\cite{Savini2012},\cite{Pisano2013}). 

The additive nature of the focusing power of flat lenses in the design of an optical system allowed us to convert an original telescope design \cite{Hargrave2014}: a telecentric refractive telescope designed to work in the frequency ranges useful for CMB experiments and Earth Observing, to a set of easy-to-manufacture thin flat lenses based on the Wood design\cite{Wood1905} and to obtain comparable performance in the same frequency range, on and off axis.

The lenses which constitute this flat-lens telescope are not only designed to have inherent low reflection loss, by having the anti-reflection coating \cite{Zhang2009} designed within the lens, but also perform a significant amount of the low-pass frequency filtering required as it is in the nature of the capacitive meshes adopted. 
Their flat nature and absence of glued coatings makes them less prone to delamination in vacuum or cryogenic conditions and provide ease of mounting in telescope supports.

In this paper, we illustrate in Section 2 the optical system which we aimed to design and its requirements. We briefly showcase, in Section 3, the advantages and limitations of the technology used in order to comprehend the manufacturing constraints which are tied to practical aspects of development time and feasibility. In Section 4 we used such constraints to define the building-block lens of the telescope via optical ray-tracing tool and conventional analytical Transmission Line modelling (TLM) to model the capacitive filtering nature of the lens structure described. This includes an iteration performed on the lens design, as the capacitive graded coating introduces further gradients by modifying the lens core thickness. 
In section 5 we validate our design and build of the single lens and the overall telescope designs with the results of an extensive spectral and spatial measurement campaign.
Conclusions and future plans for this concept are outlined in the last section.

\section{Telescope Design}\label{sec2}

No two telescope concepts are identical when their design is determined by a set of science-specific requirements.
In the pursuit of designing a full meta-material telescope (MMT), our starting point was a requirement set compatible with a number of different potential science goals for which it may be used. Avoiding an excessively prescriptive recipe we attempt to address as wide an instrumentation community as possible.

We have taken inspiration, rather than adopting an exact experimental design, from two science {\it quests} identified at the start of this activity, which are quite distinct, and their ongoing representative instruments:

\begin{itemize}
    \item A mid-frequency telescope for the CMB satellite LiteBIRD\cite{Ghigna2024}: 
    One of the instrument suites of a space mission aiming to detect the primordial signature of inflation on the Cosmic Microwave Background. This is a two-lens telecentric refractor with a 300 mm aperture and a frequency range of 89 to 224 GHz at the time of this study. It is expected that this telescope will employ powered UHMW-PE (ultra-high molecular weight poly-ethylene), the mm-wave equivalent to glass lenses, each anti-reflection coated on both sides with multiple layers of polypropylene and porous PTFE (poly-tetrafluoroethylene) in order to achieve a  $>95\%$ lens optical efficiency across the entire frequency range (excluding material absorption losses) . 
    
    \item The International Sub-Millimetre Airborne Radiometer (or ISMAR)\cite{Fox2017}: 
    covering the frequencies from 118 to 874 GHz this instrument operated by the UK Met Office aims to measure ice clouds in the upper atmosphere and is also meant as an airborne demonstrator for the Ice Cloud Imager (ICI)\cite{Bergada2016}. This is a multi-receiver instrument with a dedicated HDPE lens for each frequency channel.
    
\end{itemize}

\subsection{Scientific requirements}

Combining the main features of the above concepts with technical aspects of fabrication, our resulting design prescription for this prototype was a telescope with an aperture of 200 mm and an overall frequency range coverage from 90 to 190 GHz. The average in-band transmission (excluding absorption losses) $> 97\%$ and minimum transmission of $90\%$ were imposed for each lens resulting in an overall three-lens telescope ideally operating with an average optical efficiency $> 90\%$.
This wide spectral bandpass also allows to cover the atmospheric temperature and humidity sounding lines at 118 GHz and 183 GHz respectively.

For ease of fabrication, we chose to manufacture three identical lenses. 
The thickness of a single lens is not limited in principle but is kept to approximately 3 mm due to manufacturing practices detailed in Section 3. The resulting telescope (which uses three identical flat lenses) has an additional two degrees of freedom (lens distances $d_{12}$ and $d_{23}$) in the lens placement ranging from a single stack of 3 lenses (Configuration \#1, C1 hereafter) to a variety of inter-lens spacing combinations (C2 and C3). These designs differ minimally in their performance and moderately by overall telescope size, f-number at the focal plane and focal plane curvature.
We explored a total of 3 such configurations (Fig.\ref{fig:telescope_comparisons}) to highlight variations in telescope performance.

\subsection{The flat lens design}

To design the basic lens of our system we initially adopted a classical Wood's lens\cite{Wood1905} parabolic radial Gradient Index (GrIn) profile.
This lens replaces the conventional lens curvature as a means for changing incident ray direction with a gradient in the refractive index radially across the lens. The lens was designed with standard ray-tracing tools ANSYS-Zemax Opticstudio\cite{Zemax} with the parameters mostly constrained by manufacturing aspects. 

The index gradient is achieved similarly to the build of our first GrIn lens \cite{Savini2012}. Here we retained the capacitance-to-index function dictated by grid and spacing parameters of that device.

An important constraint of this lens type is that larger curvature (thicker) classical lenses are replaced with stronger index of refraction gradients (as well as lens thickness). 
The consequence of this trade-off is that for strong gradients in large lenses, the central index of refraction required can become quite high ($\simeq 3$) for loaded polymers and, while in principle higher indices can be engineered on the same assumptions \cite{Zhang2009}, the capacitance nature of the material also lowers the frequency cut-off of the low-pass nature of the inherent capacitive filter, limiting the high frequencies at which the lens will work efficiently. 

It is important to note that in this study the lowest cut-off considered is in any case higher than the range of frequencies of interest, providing additional out-of-band frequency filtering as part of the lens design.
The thickness of a flat GrIn lens compensates for the need of a central high index (higher gradient), as it also determines the amount of curvature that rays experience. 

The equation that governs the design of a GrIn parabolic lens in its most simple form can be written as: 
\begin{align}\label{Eq.1}
 n(r) =  n_0 - \frac{r^2}{2ft} 
\end{align}
 
with $n$ the index of refraction as a function of radius, $r$ the distance from the lens centre, $n_0$ the maximum value of the index at $r=0$, $t$ the lens thickness and $f$ the focal length of the lens. 

With this alone, in principle a focusing lens can be designed for a single‐lens telescope. Constrained by maximum index achievable ($n_0$), substrate refraction index, and physical lens radius desired ($n(r_{max}) = n_{sub}$), the design will be limited in the achievable gradient for a given diameter or vice versa. 
The overall trade-off is identified in the value of the product $f\cdot t$. 
The remaining variable, thickness “t”, in principle will then be solved in Eq.\ref{Eq.1} for the desired focal length of the lens together with a maximum radius $r_{max}\simeq 120$ mm. The latter choice is to allow for a lens with a diameter of 200 mm and some margin for a slightly larger build. 

In cases of short focal length, this is not always the best solution due to large thickness flat lenses being of increased complexity to manufacture (Section \ref{sec3}). However, the additive nature of the lenses allows the stacking of multiple lenses to be indistinguishable from a single lens if uncoated and almost equivalent to a single lens if a stack of coated lenses is used. 

For our telescope design, we proceeded by splitting the optical system into three relatively thin lenses of thickness initially set at ($t\simeq 2.6$ mm), each of which provides additional focusing until the desired telescope focal length is achieved. With large thickness being one of the most critical manufacturing issues as well as the ease of reproducibility and cost limitations, three identical lenses were considered. 

In the table below, we present the nominal design parameters of the planned lens pre-build as well as the final recovered parameters obtained as a result of the experimental verification. The difference between the two being caused by a combination of uncertainties on the relation between geometrical grid parameters and the artificial dielectric optical parameters as well as variations of thickness post-build.

\begin{figure}[h]%
    \centering
    \includegraphics[width=1.0\textwidth]{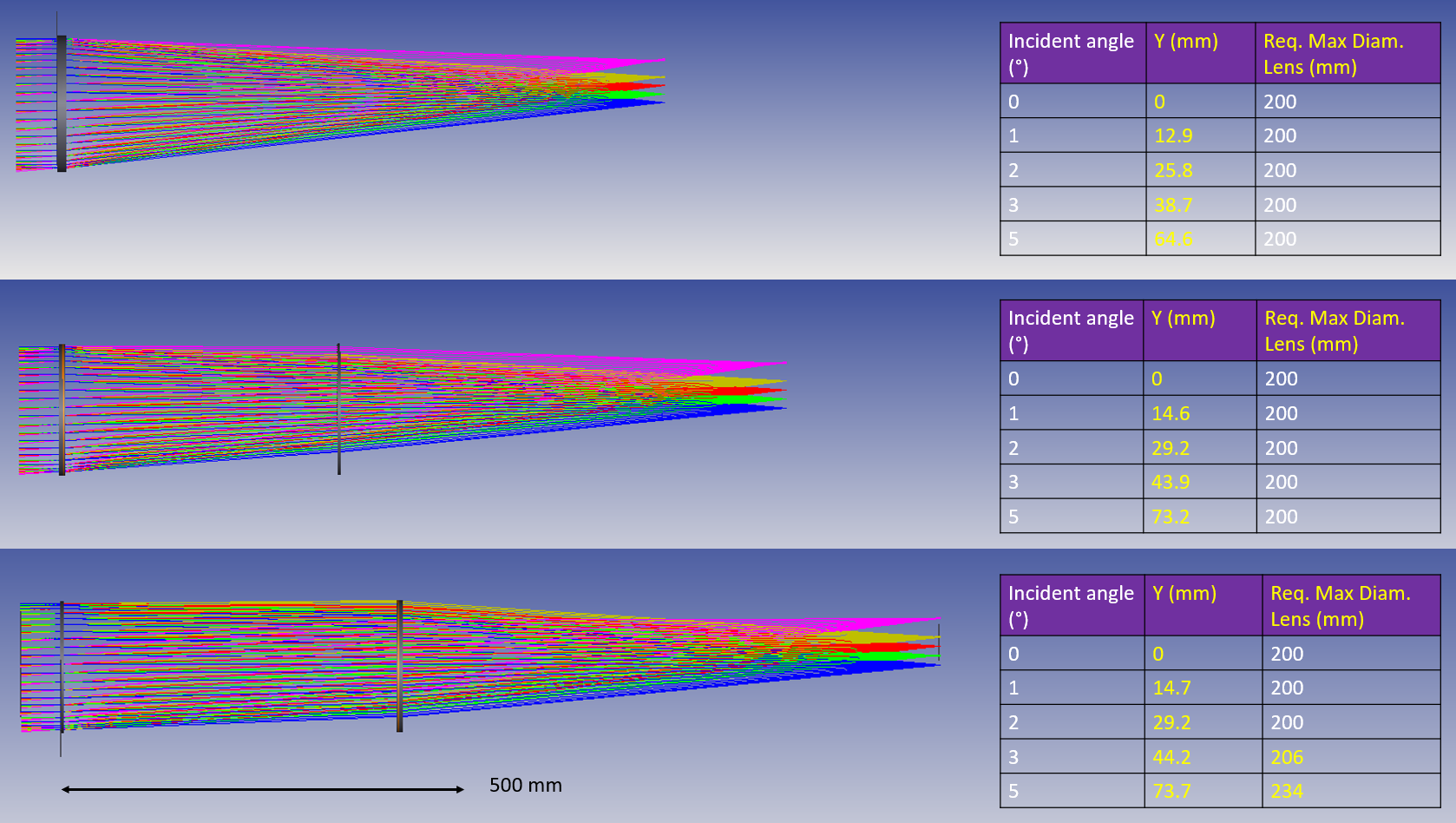}
    \caption{Three telescope configurations explored. From top to bottom: C1 - with all three lenses stacked and positioned at telescope aperture, C2 - with two lenses stacked at the aperture and one at a distance of 341.8 mm, C3 - with one lens at telescope aperture and the other two stacked at 420.3 mm.
    All configurations provide a relatively low curvature focal plane. The table on the right shows the distance from the optical axis in mm of the focus for each input angle. The right most column shows the ideal required diameter of all lenses to avoid vignetting.}
    \label{fig:telescope_comparisons}
    \end{figure}

The telescope design configurations are presented in Fig.\ref{fig:telescope_comparisons}. These designs explore the variations in lens diameter and performance for a field of view up to $\pm$5$^{\circ}$. 
The single lens in question was designed with the following parameters:

    \begin{table}[h]%
    \centering
    \caption{Lens GrIn design (uncoated)}\label{tab1}%
    \begin{tabular}{@{}llllllll@{}}
    \toprule
     Single Lens & $d^2n/dr^2$  & Thickness & Max index & n\_substrate \\
     & &  t (mm) & n\_max & & \\
    \midrule
    Nominal (pre-build) & -1.05e-4   & 2.600  & 3.000 & 1.515 &   \\
    \midrule
    Measured (post-build) & -0.94e-4  & 2.325  & 2.738 & 1.515 &  \\
    \bottomrule
    \end{tabular}\label{GrInParameters}
    \end{table}
    
    These three lenses are then considered in three different arrangements to focus on a relatively large focal plane (comparable in size to the lenses).
    These designs produce a focal plane with a f-number in the range 3 to 4.
    For the characterization of detailed low-level side-lobes and other finer features of the telescope beam, design by geometrical ray-tracing is no longer sufficient, and physical optics simulations will be required for future such design refinements. 

    Due to the distributed and relatively high values of index, the lens on its own would suffer substantially from reflection losses. So an appropriate anti-reflection coating (ARC) is tuned to the lens design.  
    In Section \ref{sec4}, we show how the inclusion of the ARC is performed and how the iteration on the single lens design is achieved.

\section{Metamaterial telescope}\label{sec3}

To design a telescope from the existing GrIn lens concept we added a similarly gradient-optimised anti-reflection coating, given the otherwise relatively poor  transmission performance caused by the impedance mismatch (especially at the centre of the lens where the index is highest). 
This was achieved by modelling the lens into discrete rings, each with a different index of refraction.

This number of rings, or “coronas” is just another design parameter which does not affect the design of the single mesh as the gradient of cells as a function of radius is based on the best fit of a second order polynomial function in the radii range and is therefore continuous. 
However, as for most computational exercises, the larger the number of modelling elements, the more accurate is the overall analytical fit likely to be.

\begin{figure}[h]%
    \centering
    \includegraphics[width=0.9\textwidth]{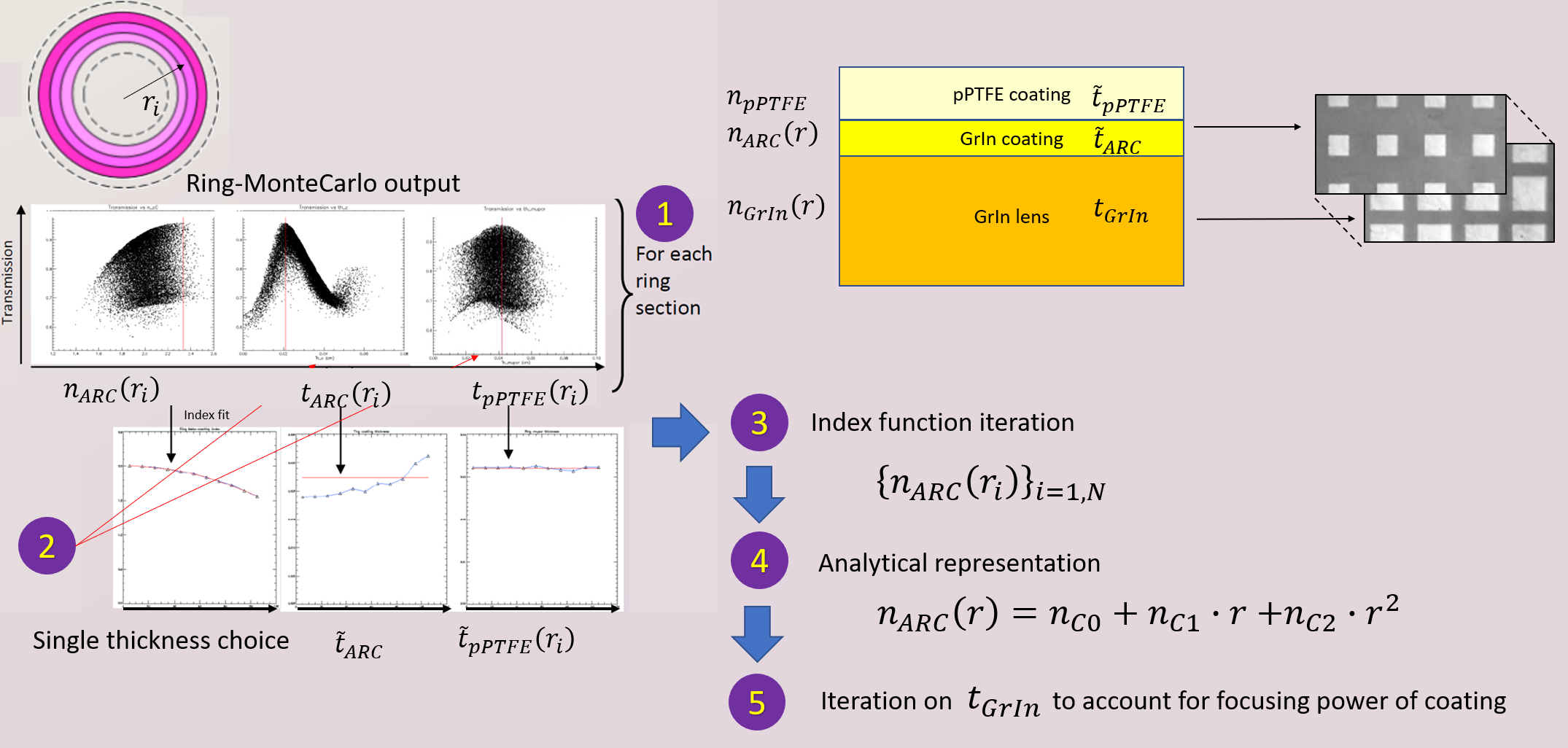}
    \caption{Starting at Top Left: In the corner is a simplistic representation of the corona sectors of the lens. Following are the steps in design iteration. 1: The result of a MonteCarlo run estimating the Transmission for the i-th ring given varying values of: ARC layer index, it's thickness and that of the porous PTFE. 2:  Due to the adoption of a one-thickness ARC layer, a single thickness of coating needs to be adopted for the entire lens (and one for the pPTFE). This is done by weighting the transmissions with the overall area of the sectors. 3: The optimized index for each sector is recalculated given the fixed thickness. 4: The curve of index values is then fitted with a second order polynomial to allow point-per-point definition of each cell index. 5: These values are used in a purposely introduced extra Gradient-1 layers in the Zemax lens parameters and the core lens thickness then adjusted to take into account the added curvature of rays imparted by the two (above and below) coating layers. Top Right: A sketch showing correspondence of patterned layers.}
    \label{fig:3-1_ARC_design}
    \end{figure}

For speed of convergence and to perform a sufficient number of exploratory simulations, a number $N_r = 30$ of rings was chosen to design the ideal ARC intermediate layer that would match the impedance between the GrIn lens core profile (in  Table \ref{GrInParameters}) and the external coating layer of pPTFE. The latter has a refractive index of approximately n = 1.23 which greatly facilitates the coating design process, providing additional tapering to the impedance from vacuum to the lens substrate material. 
The result of this analysis (summarized visually in Fig.\ref{fig:3-1_ARC_design}) leads to a second order analytical profile described by the function: 
    \begin{align}
    n_{ARC} (r)=  n_{c0} + n_{c1}\cdot r + n_{c2}\cdot r^2 
    \end{align}
    where the values for these parameters are given in Table \ref{ARCParameters} below together with the thickness of the layer:
    
    \begin{table}[h]%
    \centering
    \caption{Parametric coefficients of Gradient Index coating layers. The first row is the optimal design suggested by MC run results. The second row is the post-build best fit, but the significant difference includes an intermediate step of recalculation due to considering a thicker ($300 \mu m$) ARC-layer (as explained in Section \ref{sec4}) and resulting post-built contraction. }\label{ARCParameters}%
    \begin{tabular}{@{}llllllll@{}}
    \toprule
 $ MM_{ARC}$ & $n_{c0}$  & $n_{c1}$  & $n_{c2}$  & $t_{arc} (mm)$  \\
    \midrule
    Pre-build design & 1.923 & 1.25e-3 & -4.24e-5 & 0.220 \\
    \midrule
    Post-build recovered & 1.950 & --- & -1.44e-5 & 0.279 \\
    \bottomrule
    \end{tabular}
    \end{table}

The values for the indices were obtained by MonteCarlo (MC) simulations and the overall transmission of the pre-built {\it pPTFE-ARC-GrIn-ARC-pPTFE} stack was calculated analytically for each ring via a simple transmission line model (TLM) calculation.
The scheme and results can be seen in Fig.\ref{fig:3-1_ARC_design}.

\begin{figure}[h]%
    \centering
    \includegraphics[width=0.95\textwidth]{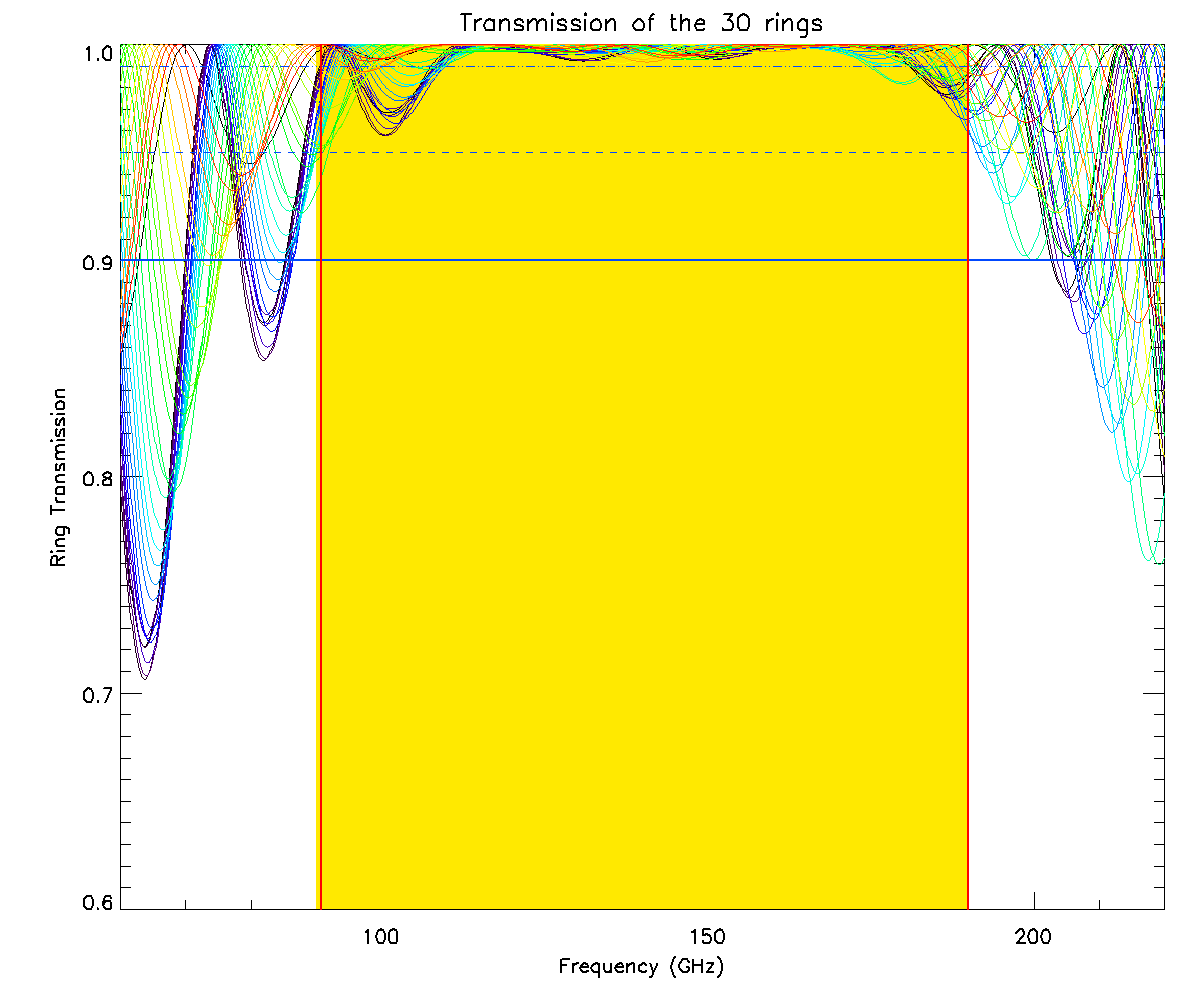}
    \caption{Modelled spectral transmission of the 30 rings are shown overlayed to the frequency band of interest between 90 and 190 GHz. A minimum in-band transmission of 0.9 for any ring was set. Values of 0.90, 0.95 and 0.99 are also shown as horizontal lines for reference. Note that this TLM does not include losses which are unavoidably present in the substrate material but considerably less than those of a thicker UHMW-PE lens. In band lossless transmission by design is shown to exceed $99\%$ in average.}
    \label{fig:3-2_v3}
    \end{figure}

The expected spectral performance (shown in Fig.\ref{fig:3-2_v3}) is the combination of each ring's transmission having been weighted for its total area. This is with an assumption of uniform intensity distribution across the lens although depending on the receiver used, it is possible that this weighting would need to be revised for further optimization. 

The next step of the design process, after the introduction of what is effectively a gradient index additional layer on each side increasing the lens thickness by almost 20\% (albeit with a lesser gradient) is to then modify (reduce) the thickness of the core part of the lens in order to maintain the desired lens convergence.
This is indeed not strictly necessary but is done in an attempt to retain as close to the original intended performance of the lens.

The final lens core thickness was then reduced to $t_{core} = 2.5$ mm and assumed to be (due to a $7\%$ overall contraction post-built) $t_{core} \simeq 2.325$.

\section{Technology and Constraints}\label{sec4}
  
The theoretical underpinning of the capacitive meshes and their filtering capabilities can be traced back to microwave theory with \cite{Marcuvitz}, evolving in the '60s and '70s \cite{Ulrich67} and \cite{Rawcliffe67} with many more seminal papers to follow. The most common application of such theory in this community to this day has been the optical and thermal filtering of mm- and submm-wavelengths in many astronomical instruments both on the ground and in space\cite{Ade2006}.

In the last two decades, it became clear that these same structures could be used to perform additional manipulation of electromagnetic waves such as phase shifting or delay for use in waveplates \cite{Pisano2008}, \cite{Zhang2011}, impedance matching for anti-reflection coatings \cite{Zhang2009}, and lenses \cite{Savini2012}, \cite{Pisano2013} and more complex concepts thereafter \cite{Moseley2017}, \cite{Pisano2023}. 

The success of these different concepts lies with the capability of depositing thin layers of copper patterns on thin polymer substrates and subsequently stacking and hot pressing these to create thicker (1 to 10 mm) sturdy structures which can then be cut to a shape to suit the instrument needs.

In the case of lenses (which are relevant to this paper), the copper structures are square periodic cells (with a period $g\sim 200\ \mu m$ in our specific case and generally substantially smaller than the wavelengths of interest). The size of the squares within the cells is the driving parameter for the cell capacitance and consequently the effective refractive index of the artificial dielectric obtained by stacking multiple layers.  

The clean-room processes which have been improved through the decades are still tied to manufacturing details which have had to (for different experimental needs), each time, undergo upgrades or enhancements, such as evaporation and etching facilities, but are still limited by technical aspects of the facilities used. As such, while extremely large devices ($>1m$) are theoretically possible, and have indeed been achieved (\cite{Pisano2018},\cite{
Braithwaite}), moderate sizes ($10$ to $30$ cm) are preferable when requiring to produce tens or hundreds of layer structures. The process of grid alignment, especially for vast number of layers can be complex, so employing a technique such as that of effective artificial dielectric materials
which is more forgiving in the overall alignment tolerances is tempting.

The details of mesh or pattern geometry correspondence with effective refractive index can be found in a previous work \cite{Savini2012} as they have not substantially changed. In order to limit the natural cut-off of the capacitance structure, a maximum index of $n\simeq 3.0$ was initially identified, but resulted in a post-build reduced value of $n = 2.738$.

We additionally performed finite element analysis using ANSYS High Frequency Simulations [HFSS] to verify that multiple cell mis-alignments would not compromise substantially our design. This was done by monitoring the effective refractive index behaviour by fitting the Fabry-Perot etalon function of an equivalent thickness as similar single cells were mis-aligned, randomly first and purposely by half a period in a worst case example. The index of refraction variation was observed to be a maximum of $\Delta n \simeq 1\%$.

Similarly, the variety of measured values for the refractive indices of different polymer substrates for different suppliers was initially a matter of concern, but similar simulations adopting the expected range in known indices for different suppliers of polymer substrate 
showed limited impact on the recovered effective index value, well within the previous range of parameters.  

The overall diameter of the lens is subject to the strongest limitation so far of this design. Having fixed a maximum value for $n_0$, the maximum index gradient available is set by the difference of the central index to that of the substrate (reached at the edge of the lens). This is equivalent in a sense to how much curvature and therefore convergence can be imparted on the incoming rays. 
It has been possible to design and build \cite{Moseley2018} larger Fresnel-like lenses with the same technology. This could potentially offer a solution with the added complexity of combining Fresnel lenses for future concepts.

For this telescope concept we chose an allowed size of $200$ mm diameter. 
This is both a standard size (for the manufacturing) of previously built filters and allowed for the telescopes shown in Fig.\ref{fig:telescope_comparisons} to be achieved with overall lengths comparable to that of the concepts we were inspired by. 

The first of two other practical limitations on the design is the commercial availability of set thicknesses of pPTFE. While the ideal design had a thickness of $\sim 350 \mu m$, the only thickness available to us at the time was $ 450 \mu m$.
The second limitation (which can be amended in future concepts) is the discrete thickness steps for both core- and ARC-grin thicknesses.
The effective index material is based on multi-grid simulation of a number of grids $N\ge3$.
However, having simulated grids at a set grid-distance of $t=100\mu m$, this provides the discrete step of thickness for this design.
While the core part of the lens was set to $d=2.5mm$ after the inclusion of the ARC, the optimum thickness of ARC was instead increased to $300 \mu m$
with the likely effect of a small shift of the coating effectiveness towards longer wavelengths.

The final manufactured recipe is listed in Table \ref{tab3}.

    \begin{table}[h]%
    \centering
    \caption{Final recipe for single GrIn lens. Post-build an overall decrease in total thickness of $7\%$ was observed. While exact partition of such decrease among components could not be verified without damaging the lens, this quantity was consistent with other previously built components with the same technology (with and without coating). Measurement uncertainties can be all considered at the $5\mu m$ level.}\label{tab3}%
    \begin{tabular}{@{}llllll@{}}
    \toprule
      & $t_{pPTFE} (mm)$  & $t_{ARC} (mm)$  & $t_{core} (mm)$  \\
    \midrule
    Recipe & 0.450 & 0.300 & 2.500 \\
     \midrule
    Post-build & 0.418 & 0.279 & 2.325 \\
    \bottomrule
    \end{tabular}
    \end{table}

\section{Measurements and Discussion}

The MMT measurement campaign can be broadly defined in three phases: 
First, a spectral test of the ARC-integrated lens performed with a custom-designed Fourier Transform Spectrometer (FTS) the
details of which can be found in previous similarly performed tests \cite{Zhang2011}, \cite{Moseley2017}. The second measurement performed was to check summarily the beam properties of the single lens and although a similar lens had previously been measured\cite{Savini2012}, the changes in parameters and addition of ARC warranted verification for the purpose of telescope design.
The final set of measurements saw the three identical lenses combined as discussed in section \ref{sec2} to create three different telescope configurations for which we performed near-field beam scans (intended as orthogonal ``cuts") across the focal plane. 
The choice of this particular measurement, given the setup shown in Fig.\ref{Fig:5-2AB}, is favoured with respect to beam patterns obtained via rotation of the telescope about the vertical axis through the centre of the telescope aperture. This is mostly due to the impracticality of rotating a set of telescope configurations the length of which can vary from $\sim 0.7-1.1m$ with a VNA as a receiver.

\subsection{Spectral performance of the anti-reflection coated lens }

The output of the FTS was throughput-matched to a pair of ultra-high molecular weight polyethylene (UHMW-PE) fast (low $F$-number) lenses to form a collimated beam testing range with a smaller diameter than that of the lenses we aimed to test.
The second lens focused to a cooled bolometer detector operating at a temperature of 4 K. 

This allowed us to spectrally test small portions of the lens at a time, given the expected variable spectral behaviour of different capacitive portions of the lens.

As positioning the single lens in the collimated beam range would also alter the focusing of this spectral test range, the position of the second PE lens was adjusted according to predicted lens performance to maximize transmission at the detector. This was empirically verified to be the case
by stepping the lens manually in small increments along the optical axis and observing the overall signal strength at the detector.

Note that the lens was designed to be of a much longer focal length than those employed in this testbed, and therefore the alteration to the testbed focus is relatively small and the shift of the detecting apparatus easy to impart. 

\begin{figure}[h]%
\centering
\includegraphics[width=0.9\textwidth]{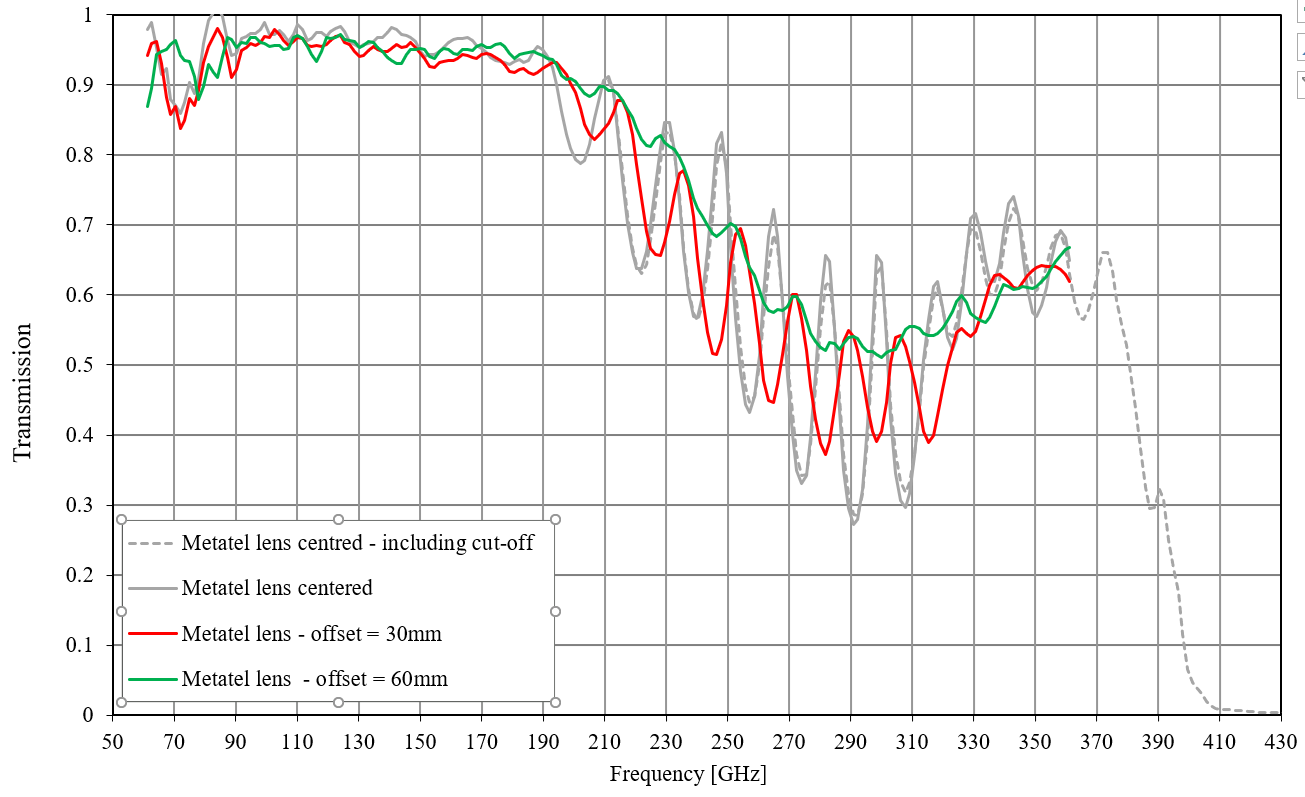}
\caption{Measured transmission of the lens on a Fourier Transform Spectrometer. Separate measurements were performed using a small circular aperture to compare average spectral performance of regions progressively offset from the centre. 
For the centre region, an initial full range spectrum including the capacitive cut-off is included (dashed-grey). A subsequent set of three measurements done on three different parts of the lens show respectively: grey - the centre of the lens; red and green - offset measurement (as per values in the plot legend).}
\label{fig:5-1_spectral_test}
\end{figure}

Measurements were made through the central portion of the lens, and at two off-centre positions to check for variations in spectral response due to the different overall values of index and its local gradient in these regions and with knowledge of the expected coating designed. 
A 20 mm diameter aperture was used to test different portions of the lens as this was offset with respect to the optical axis. 
The average performance in Fig.\ref{fig:5-1_spectral_test} shows a minimum transmission well above 90\% across the range 90-195 GHz which includes absorption losses.

The fringing occurring at higher frequencies outside of the coated-optimized frequency region is as expected more pronounced in the spatially central region of the lens, where a quasi-constant index value is found across the aperture used and becomes progressively smoothed as the variation of index of the lens within the measurement aperture is such that the combined Fabry-Perot fringes average out.

The average transmission of the single lens in the adopted frequency range, for the three separate regions of the lens is measured with an uncertainty of $\sim 1\%$ at $ 96\%, 95\% $ and $ 95\% $ respectively (including absorption losses).

\subsection{Beam cut of single lens }

\begin{figure}[h]%
\centering
\includegraphics[width=0.48\textwidth]{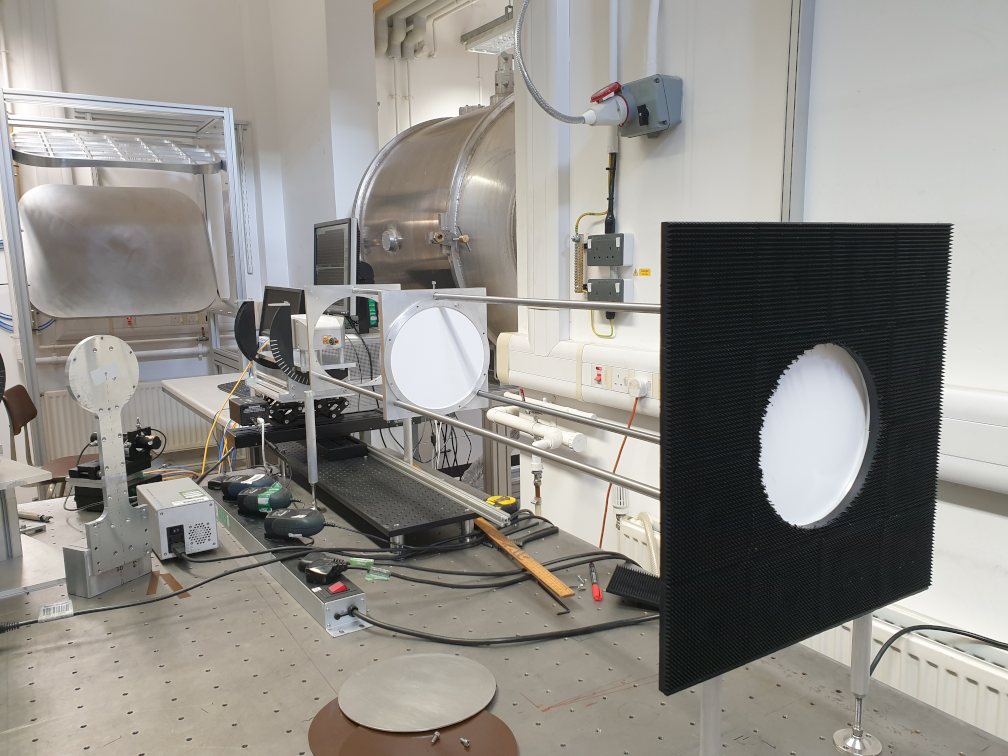}
\includegraphics[width=0.48\textwidth]{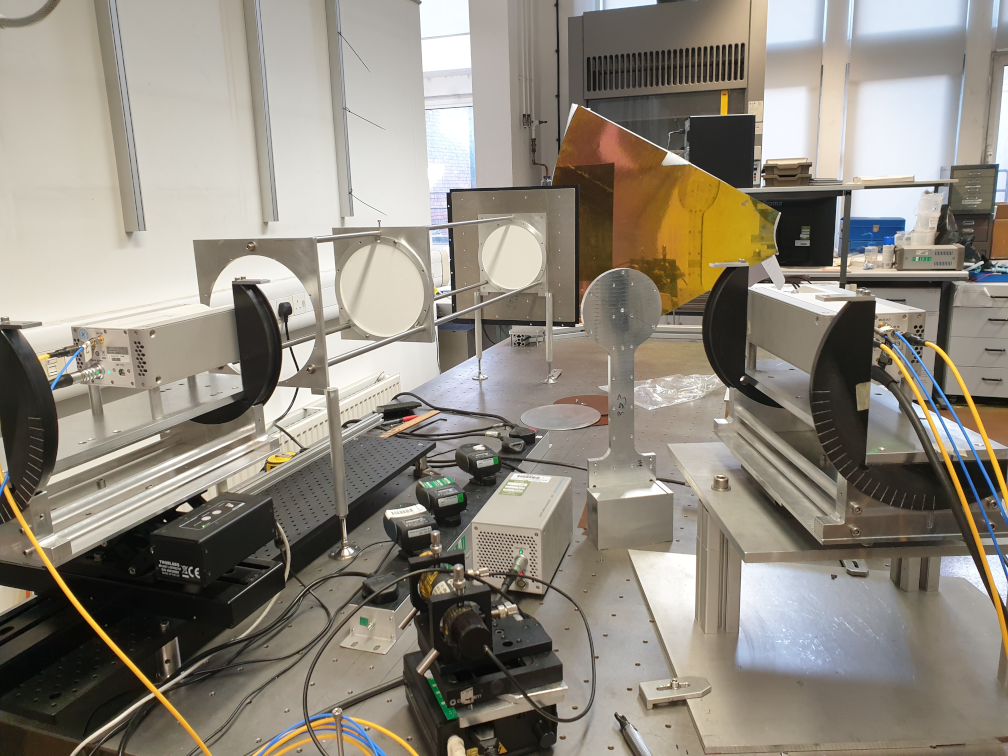}
\caption{The test setup can be seen (Left) looking towards the receiver from the side of the collimating mirror and (Right) looking towards the collimating mirror from an intermediate angle behind both launcher and receiver.}
\label{Fig:5-2AB}
\end{figure}

All spatial beam-cuts were performed with a Rhode \& Schwarz Z67 vector-network analyser (VNA) with ZC170 millimetre-wave frequency converters (110-170 GHz) using standard rectangular horn gain antennas (Model Flann 29240-20) as launcher and receiver. The collimated beam directed toward the lens is obtained using a sector of a large (2m diameter), (carbon fibre reinforced polymer) CFRP gold coated mirror. The setup is shown in Fig.\ref{Fig:5-2AB}.
The receiver horn is mounted on a three-axis motorized stage allowing horizontal and vertical scans across the beam as well as through-focus measurements. The frequencies spanned with the VNA range from 110 to 170 GHz. 

Fig.\ref{Fig:5-2AB} shows the large cage mounts that allow the testing of up to three lenses (when all three are kept separate), but in the single lens test, only one mount for a single lens was used.
The front (aluminium) mount used is covered with absorbing tiles \cite{TKtiles}.
In all following measurement descriptions, the reference frame convention adopted is ``z" for the optical axis and ``x" for the horizontal beam-cut scanning direction.

While the test setup was designed to perform tests of the overall MMT, an attempt to a preliminary beam cut of the single lens was performed to verify our expected lens behaviour (similarly to the spectroscopic tests), despite the expected single lens focal length exceeded slightly the maximum available lens-positioning range of our test setup.

Our estimated post-build single lens focal length $f = 2250$ mm was longer than the maximum available testbed receiver-lens distance (2080 mm)\footnote{The test-bed was designed with the testing of the full telescope in mind requiring overall substantially less optical axis length} so the latter was adopted for this scan. 

\begin{figure}[h]%
\centering
\includegraphics[width=0.8\textwidth]{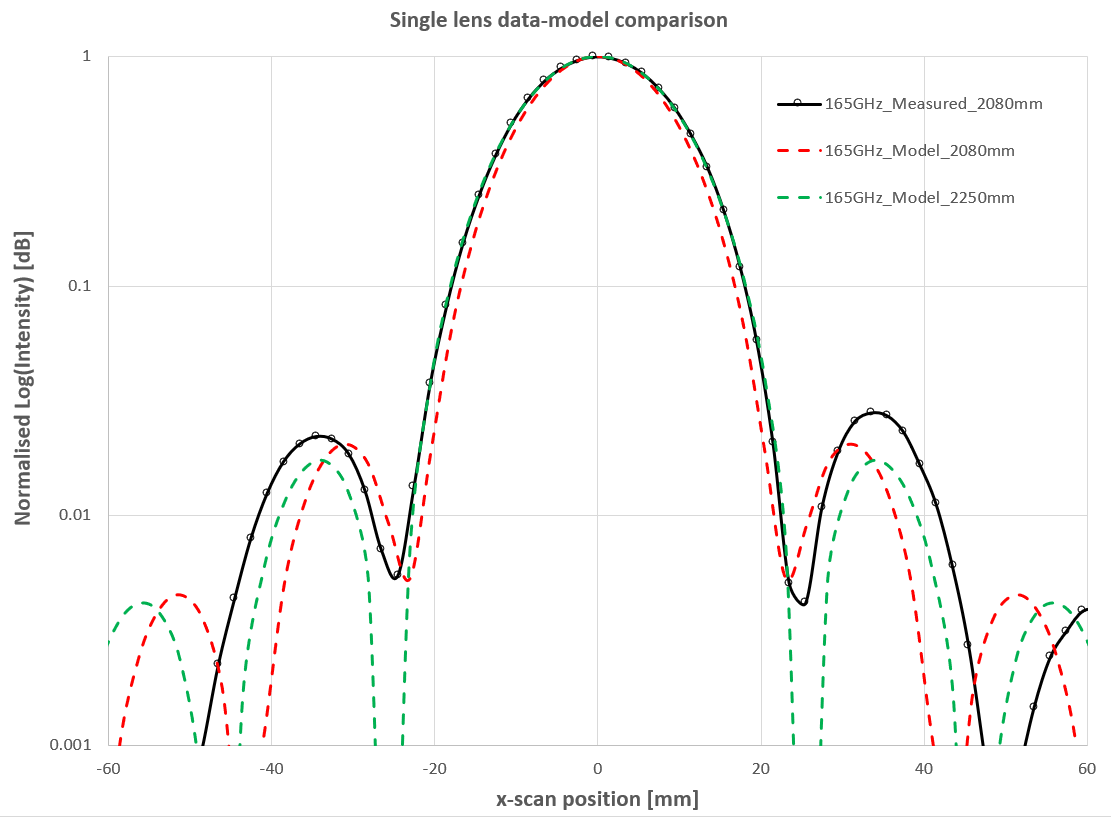} 
\caption{The data (black) shows a slightly larger beam cut than the equivalent beam Huygens profile at the same distance (red dotted). The half-width of the off-focus beam (from peak to first zero) is $1$ mm ($\sim 4\%$) larger than expected. However, given the mentioned approximations in the ``contracted" post-build and the uncertainties on the index determination, this is considered to be an acceptable discrepancy. A model beam cut for a longer (2250 mm) focal distance is shown for comparison (green dotted).}\label{Fig:singlelensbeam} 
\end{figure}

A comparison between measurement and model is shown for $\nu = 165$ GHz at that particular distance in Fig.\ref{Fig:singlelensbeam}.
The data measured are shown as a black curve, while for the lens recipe two model beam cuts are shown at the same distance (green and red) respectively for the nominal thicknesses recipe modified for a $\sim 7\%$ compression of the overall lens thickness observed after fabrication. 

\subsection{Meta-material Telescope beam cross section.}

The same measurements as described above were performed by placing a lens in
each of the three movable frames in the optical cage visible in Fig.\ref{Fig:5-2AB}.
Best focus for each configuration was determined by strength of signal and low-level side-lobes. Through focus measurements were performed with beam cuts at different z-positions to assess the focus. In this section, the measurements are compared with modelled physical beam cuts both on- and off-axis. 
As mentioned in section \ref{sec2}, we performed measurements in three configurations to confirm that the performance of the lenses and their combinations is well understood and reproducible. The three configurations are defined in Table \ref{Table:3-lenses}.

\begin{table}[h]%
\centering
\caption{Telescope Configurations by design. \\ {\small Distances are quoted to the closest mm. (Lens thicknesses of 4 mm each included in overall telescope length)}}\label{TelescopeConfigs}%
\begin{tabular}{@{}c|cccc@{}}
\toprule
Configuration & $d_{12}$  & $d_{23}$   & Focus & Overall telescope \\
 \# & (mm) & (mm)  & (mm) & length (mm) \\
\midrule
1 & 0 & 0 & 750 & 762\\
2 & 0 & 342 & 599 & 953\\
3 & 420 & 0 & 716 & 1148\\
\bottomrule
\end{tabular}\label{Table:3-lenses}
\end{table}

\begin{figure}[h]%
\centering
\includegraphics[width=0.7\textwidth]{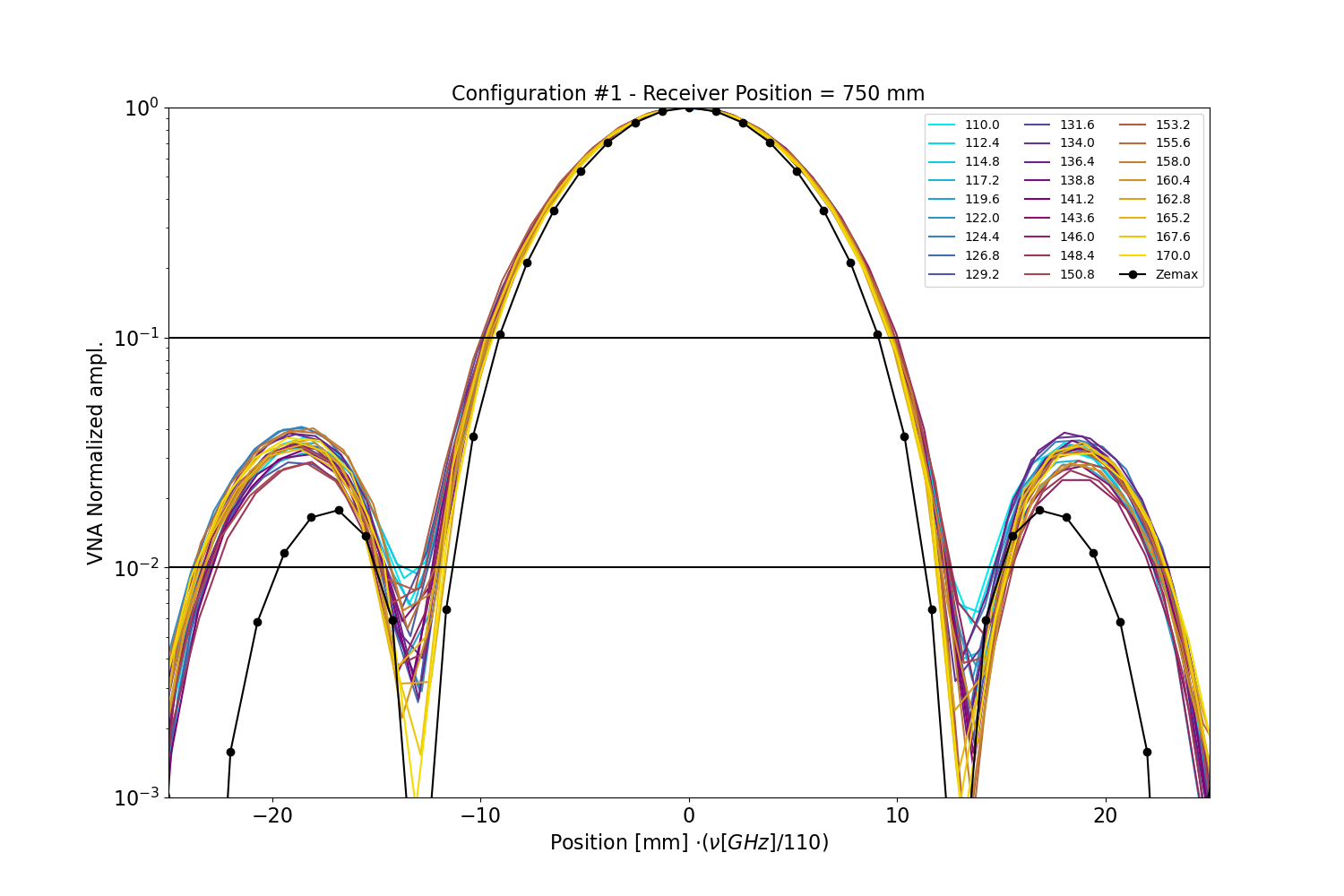} 
\includegraphics[width=0.7\textwidth]{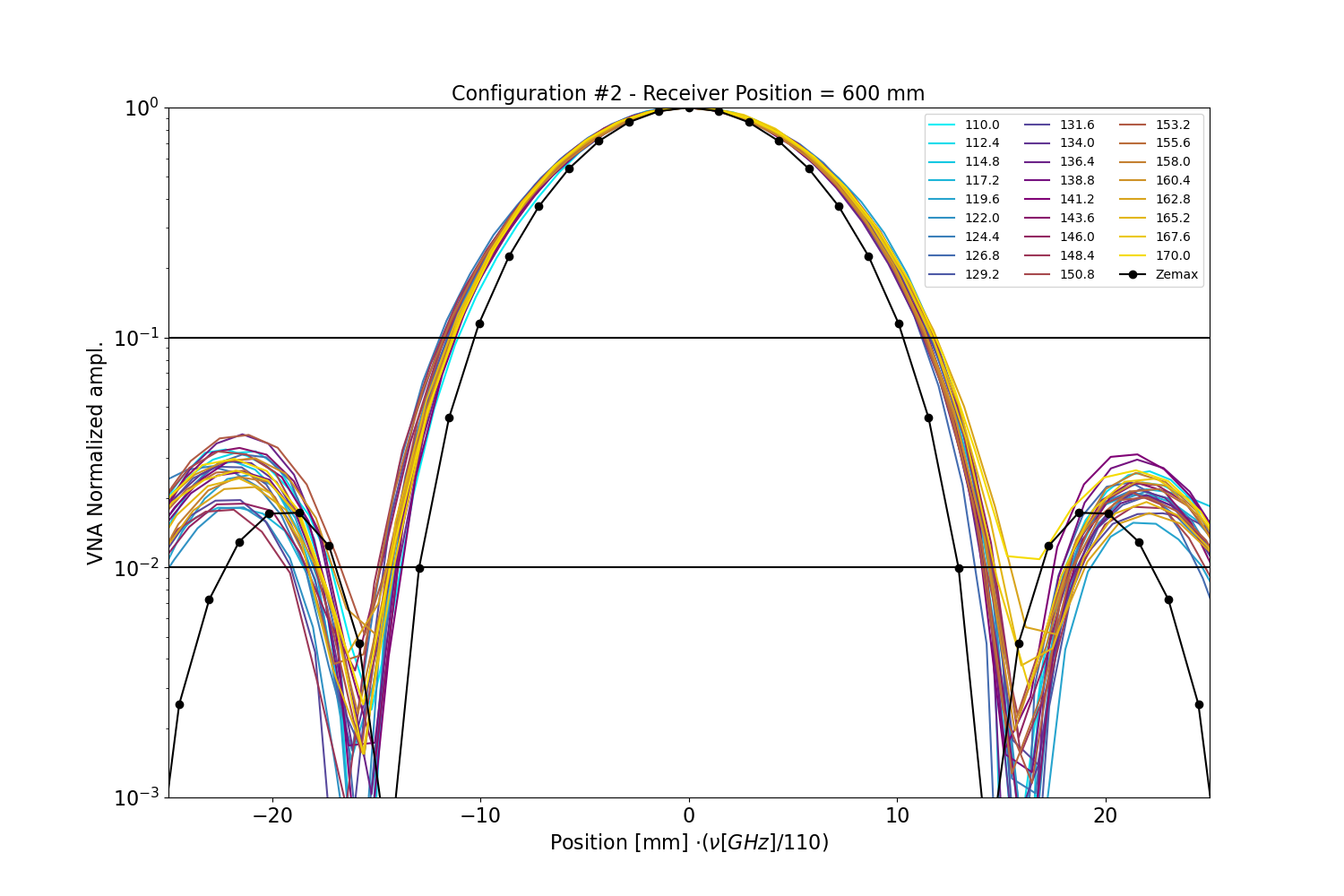} 
\includegraphics[width=0.7\textwidth]{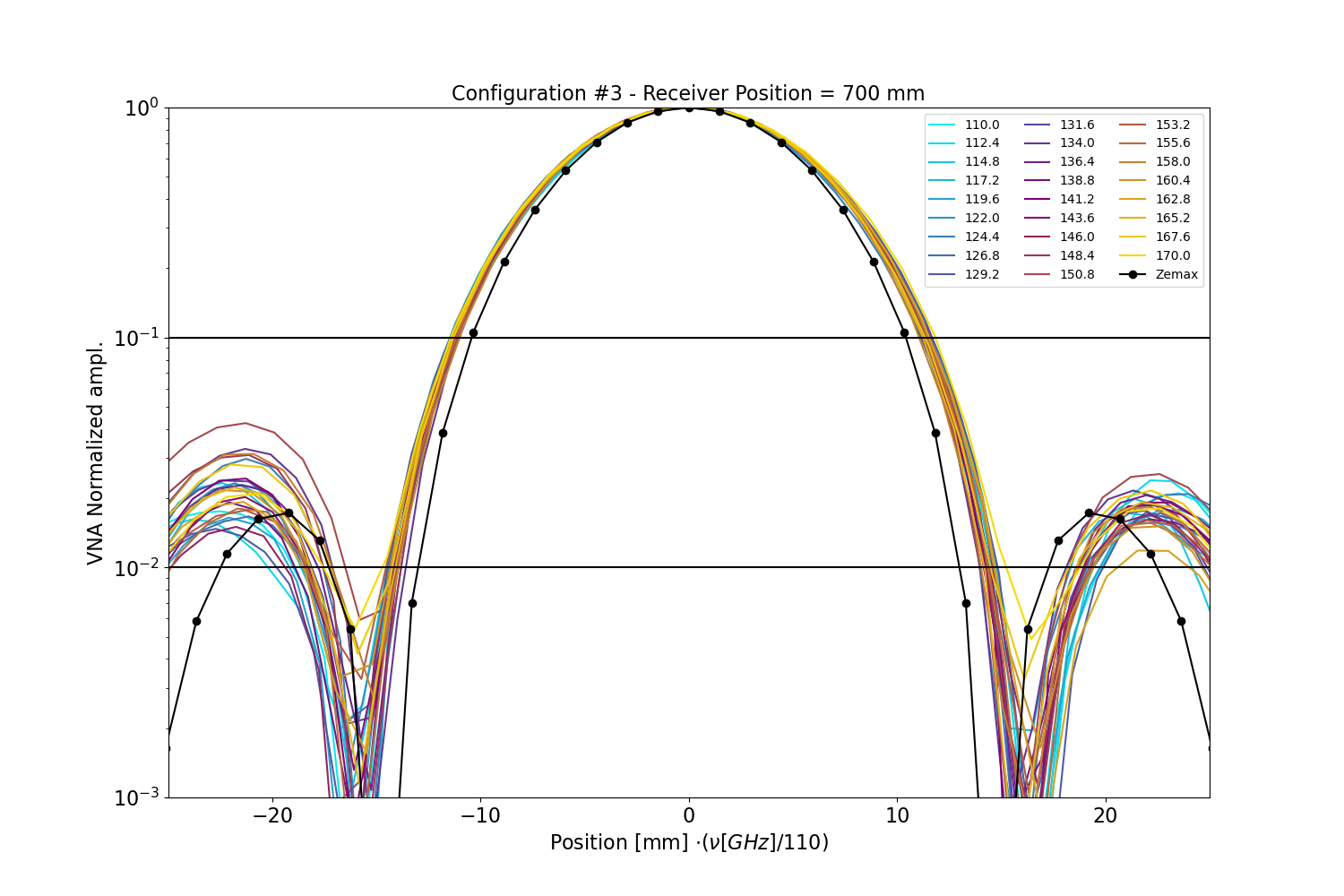} 
\caption{Beams for 26 different frequencies in a range 110-170 GHz, with the x-axis scaled with wavelength for each beam and compared for each configuration to the equivalent expected theoretical Huygens PSF cross section (Zemax).}
\label{Fig:5-4_MMT_beamcuts} 
\end{figure}

In Fig.\ref{Fig:5-4_MMT_beamcuts} we show the beam cuts measured at 26 frequencies spanning from 110 to 170 GHz taken each time to the closest position to the expected focal distance and compared to the modelled focus with a physical Huygens PSF cross-section beam cut at the same distance as that of the measurement.

The beam cuts appear consistent with the expected scaling with wavelength in the diffraction regime (thus each frequency data set in Fig.\ref{Fig:5-4_MMT_beamcuts} is plotted with respect to the product $\lambda \cdot x$).

\subsection{Off-axis properties}\label{offaxis}

The on-axis behaviour of the lens is consistent with the expected model. To make sure that the lens performs accordingly as a GrIn lens off axis, we also compared measurements and model for rays up to the maximum available off-axis angle allowed ($\pm 1.5^{\circ}$) due to the combination of collimator focal length and launcher scanning stage.

\begin{figure}[h]%
\centering
\includegraphics[width=0.9\textwidth]{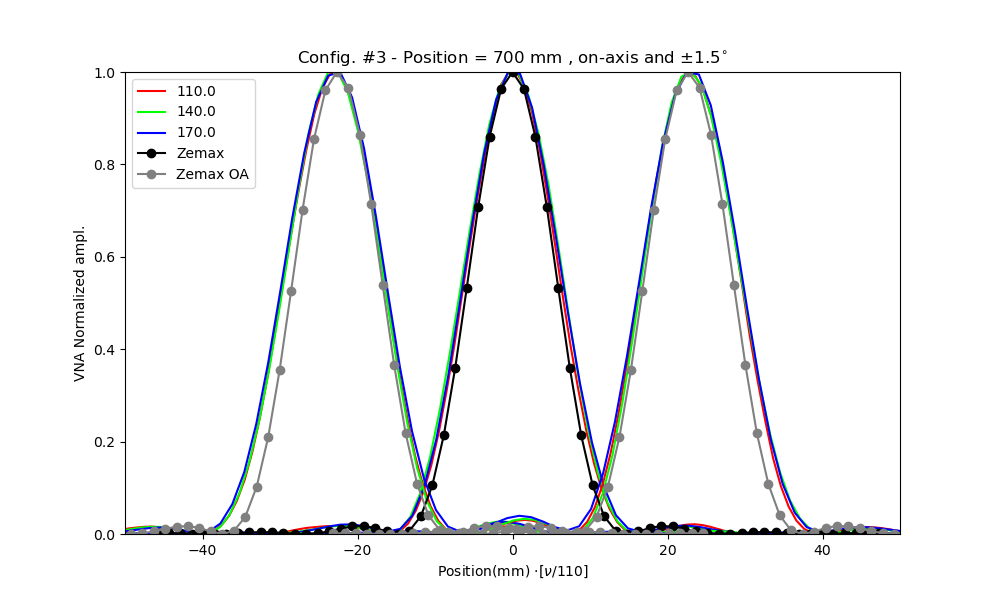} 
\caption{The beam cut comparison performed at frequencies 110,140 and 170 GHz and at angles $-1.5^{\circ}$,$0^{\circ}$,$-1.5^{\circ}$. Circle marker-lines show the respective expected on-axis (black) and off-axis (gray) profile from optical ray-tracing simulations.}
\label{Fig:5-MMT_offaxis} 
\end{figure}

The comparison with the model suggests either a small additional uncertainty in the performance of the lens off-axis degenerate with an angle uncertainty of $\sim 0.2^{\circ}$.

The measurements were performed in an identical way to the previous on-axis measurement. A subset of these for configuration $\# 3$ is shown in Fig. \ref{Fig:5-MMT_offaxis} exhibiting no appreciable distortion of the beam for the small angles in question.

\begin{figure}[h]%
\centering
\includegraphics[width=0.9\textwidth]{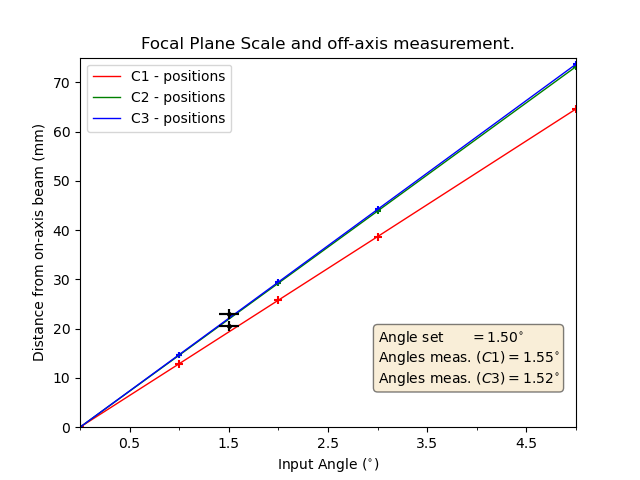}
\caption{Telescope plate scale estimated for the three configurations compared with the measured case. Measured off-axis focus is shown for two configurations as the black dots with relevant error bars, showing good agreement with the expected focal plane scale.}
\label{Fig:5-6_fps} 
\end{figure}

The lens was modeled with the post-build parameters. Fig.\ref{Fig:5-6_fps} shows the expected focus distance on the focal plane with angle for the three different configurations. 

As the lens modelling and measurement seem to compare quite well, we estimate the focal plane curvature of this system, by the use of our model.

Maximum variation of focal distance off axis at $5^{\circ}$ angle is estimated for the three MMT configurations to be respectively $[0.0037\%,0.08\%,0.18\%]$ suggesting that these lenses can indeed be used (at least in this angle range) for an extended focal plane array (rather than only in a single pixel or receiver configuration).

\section{Conclusions}

We have shown the end-to-end process of design, fabrication, and measurements of a first mm-wave full optical system incorporating only meta-material flat lenses.
The flat polymer lenses are built with a refractive index engineering method based on photo-lithography and polymer heat-bonding. This metamaterial technology allows us to replicate the performance of a standard mm-wave telescope (compared with a diameter-scaled equivalent concept \cite{Hargrave2014}) with lenses of a fraction of their thickness (one polymer lens of $\sim 1$ cm compared to a set of lenses with a total central thickness of $\sim 7$ cm) and $\sim 20\% $ of the mass. The latter mass reduction is by direct polymer weight comparison, with the metamaterial lens including an embedded ARC as well as out-of-band filtering and reduced mount complexity, the mass savings are likely to be much higher.
The frequency broad-band performance of these lenses is compliant to that of a range of astrophysical and Earth Observing telescopes.

The overall in-band optical efficiency of the entire telescope cannot be easily measured in the same way as the single lens, so the third power of the measurement of one such lens is considered and expected to be $\sim (85 \pm 2)\%$.

The flat nature of the lens significantly simplifies the mount design for cryogenic telescopes and minimises any potential deformations that ensue with profiled lenses during cooling.
The measured performance of the single lens is consistent within acceptable fabrication tolerances and can be further improved with empirical knowledge after fabrication and measurement.

The main beam of a telescope comprising three such lenses has been measured in three different configurations both on- and off-axis, allowing to test differences in focal plane scale and focal plane curvature. Results compare closely in both cases allowing relatively simple first-order modelling with ray-tracing tools.

For some more complex applications such as mm-wave cosmology, knowledge of side-lobes (both near and far) can be a key design factor, and future designs (and measurement setup) will need to include such requirements and constraints in both the design process and measurement thresholds.

The full meta-material telescope has been achieved through a design iteration and under substantial constraints (having opted to design three identical lenses). 
A full-scale future telescope for a dedicated experiment would likely allow each lens to be tuned separately to further improve the overall telescope properties.



\bigskip








\backmatter



\section*{Funding} 
The authors would like to acknowledge the funding supporting this activity from the Centre for Earth Observing Instrumentation of the UK Space Agency (CEOI-UKSA) contract number RP10G0435A203, ``Ultra-lightweight metamaterial optics for Earth Observation applications (Meta-Tel)"


\noindent

\bigskip

\bibliography{reference}


\begin{thebibliography}{30}
\ifx \bisbn   \undefined \def \bisbn  #1{ISBN #1}\fi
\ifx \binits  \undefined \def \binits#1{#1}\fi
\ifx \bauthor  \undefined \def \bauthor#1{#1}\fi
\ifx \batitle  \undefined \def \batitle#1{#1}\fi
\ifx \bjtitle  \undefined \def \bjtitle#1{#1}\fi
\ifx \bvolume  \undefined \def \bvolume#1{\textbf{#1}}\fi
\ifx \byear  \undefined \def \byear#1{#1}\fi
\ifx \bissue  \undefined \def \bissue#1{#1}\fi
\ifx \bfpage  \undefined \def \bfpage#1{#1}\fi
\ifx \blpage  \undefined \def \blpage #1{#1}\fi
\ifx \burl  \undefined \def \burl#1{\textsf{#1}}\fi
\ifx \doiurl  \undefined \def \doiurl#1{\url{https://doi.org/#1}}\fi
\ifx \betal  \undefined \def \betal{\textit{et al.}}\fi
\ifx \binstitute  \undefined \def \binstitute#1{#1}\fi
\ifx \binstitutionaled  \undefined \def \binstitutionaled#1{#1}\fi
\ifx \bctitle  \undefined \def \bctitle#1{#1}\fi
\ifx \beditor  \undefined \def \beditor#1{#1}\fi
\ifx \bpublisher  \undefined \def \bpublisher#1{#1}\fi
\ifx \bbtitle  \undefined \def \bbtitle#1{#1}\fi
\ifx \bedition  \undefined \def \bedition#1{#1}\fi
\ifx \bseriesno  \undefined \def \bseriesno#1{#1}\fi
\ifx \blocation  \undefined \def \blocation#1{#1}\fi
\ifx \bsertitle  \undefined \def \bsertitle#1{#1}\fi
\ifx \bsnm \undefined \def \bsnm#1{#1}\fi
\ifx \bsuffix \undefined \def \bsuffix#1{#1}\fi
\ifx \bparticle \undefined \def \bparticle#1{#1}\fi
\ifx \barticle \undefined \def \barticle#1{#1}\fi
\bibcommenthead
\ifx \bconfdate \undefined \def \bconfdate #1{#1}\fi
\ifx \botherref \undefined \def \botherref #1{#1}\fi
\ifx \url \undefined \def \url#1{\textsf{#1}}\fi
\ifx \bchapter \undefined \def \bchapter#1{#1}\fi
\ifx \bbook \undefined \def \bbook#1{#1}\fi
\ifx \bcomment \undefined \def \bcomment#1{#1}\fi
\ifx \oauthor \undefined \def \oauthor#1{#1}\fi
\ifx \citeauthoryear \undefined \def \citeauthoryear#1{#1}\fi
\ifx \endbibitem  \undefined \def \endbibitem {}\fi
\ifx \bconflocation  \undefined \def \bconflocation#1{#1}\fi
\ifx \arxivurl  \undefined \def \arxivurl#1{\textsf{#1}}\fi
\csname PreBibitemsHook\endcsname

\bibitem{Planck1}
\begin{barticle}
\bauthor{\bsnm{Ade}, \binits{P.A.R.}},
\bauthor{\bparticle{et} \bsnm{al.}}:
\batitle{Planck pre-launch status: The optical architecture of the hfi}.
\bjtitle{Astronomy and Astrophysics}
\bvolume{520}(\bissue{A11}),
\bfpage{7}
(\byear{2010})
\end{barticle}
\endbibitem

\bibitem{Simons2024}
\begin{barticle}
\bauthor{\bsnm{Galitzki}, \binits{N.}},
\bauthor{\bparticle{et} \bsnm{al.}}:
\batitle{The simons observatory: Design, integration, and testing of the small aperture telescopes}.
\bjtitle{The Astrophysical Journal Supplement Series}
\bvolume{274:33},
\bfpage{27}
(\byear{2024})
\end{barticle}
\endbibitem

\bibitem{Ghigna2024}
\begin{barticle}
\bauthor{\bsnm{Ghigna}, \binits{T.}},
\bauthor{\bparticle{et} \bsnm{al.}}:
\batitle{The litebird mission to explore cosmic inflation}.
\bjtitle{Space Telescopes and Instrumentation 2024: Optical, Infrared, and Millimeter Wave}
\bvolume{13092},
\bfpage{1309228}
(\byear{2024})
\end{barticle}
\endbibitem

\bibitem{Yang2011}
\begin{barticle}
\bauthor{\bsnm{Yang}, \binits{H.}},
\bauthor{\bparticle{et} \bsnm{al.}}:
\batitle{The fengyun-3 microwave radiation imager on-orbit verification}.
\bjtitle{IEEE Transactions on Geoscience and Remote Sensing}
\bvolume{49},
\bfpage{4552}--\blpage{4560}
(\byear{2011})
\end{barticle}
\endbibitem

\bibitem{Ilias2013}
\begin{barticle}
\bauthor{\bsnm{Ilias}, \binits{M.}},
\bauthor{\bsnm{Grabarnik}, \binits{S.}},
\bauthor{\bsnm{Caron}, \binits{J.}},
\bauthor{\bsnm{Bezy}, \binits{J.-L.}},
\bauthor{\bsnm{Loiselet}, \binits{M.}}:
\batitle{The metop second generation 3mi instrument}.
\bjtitle{Proc. SPIE 8889, Sensors, Systems, and Next-Generation Satellites XVII}
\bvolume{88890},
\bfpage{88890}--\blpage{113}
(\byear{2013})
\end{barticle}
\endbibitem

\bibitem{Jornet2023}
\begin{barticle}
\bauthor{\bsnm{Jornet}, \binits{J.M.}},
\bauthor{\bsnm{Knightly}, \binits{E.W.}},
\bauthor{\bsnm{Mittleman}, \binits{D.M.}}:
\batitle{Wireless communications sensing and security above 100 ghz}.
\bjtitle{Nature Communications}
\bvolume{14},
\bfpage{1}--\blpage{10}
(\byear{2023})
\end{barticle}
\endbibitem

\bibitem{ofcom2020}
\begin{botherref}
\oauthor{\bsnm{OFCOM-UK}}:
Statement: Supporting innovation in the 100-200 {GH}z range
(1st October 2020).
\url{https://www.ofcom.org.uk/__data/assets/pdf_file/0024/203829/100-ghz-statement.pdf}
Accessed 2024-01-02
\end{botherref}
\endbibitem

\bibitem{Lupi2020}
\begin{barticle}
\bauthor{\bsnm{Lupi}, \binits{t.}},
\bauthor{\bparticle{et} \bsnm{al.}}:
\batitle{Microwave imager for metop-sg: development status and instrument verification}.
\bjtitle{IEEE Proceedings of the 16th Specialist Meeting on Microwave Radiometry and Remote Sensing for the Environment}
(\byear{2020}).
\doiurl{10.1109/MicroRad49612.2020.9342547}
\end{barticle}
\endbibitem

\bibitem{Ade2006}
\begin{barticle}
\bauthor{\bsnm{Ade}, \binits{P.A.R.}},
\bauthor{\bsnm{Pisano}, \binits{G.}},
\bauthor{\bsnm{Carole}, \binits{T.}},
\bauthor{\bsnm{Weaver}, \binits{S.}}:
\batitle{A review of metal-mesh filters}.
\bjtitle{Proceedings Volume 6275, Millimeter and Submillimeter Detectors and Instrumentation for Astronomy III; 62750U (2006)}
(\byear{2006}).
\doiurl{10.1117/12.673162}
\end{barticle}
\endbibitem

\bibitem{Wood1905}
\begin{bbook}
\bauthor{\bsnm{Wood}, \binits{R.W.}}:
\bbtitle{Physical Optics}.
\bpublisher{The MacMillan Company}, \blocation{???}
(\byear{1905})
\end{bbook}
\endbibitem

\bibitem{Zhang2009}
\begin{barticle}
\bauthor{\bsnm{Zhang}, \binits{J.}},
\bauthor{\bparticle{et} \bsnm{al.}}:
\batitle{“new artificial dielectric metamaterial and its application as a terahertz antireflection coating}.
\bjtitle{Applied Optics}
\bvolume{48}(\bissue{35}),
\bfpage{6635}--\blpage{6642}
(\byear{2009})
\end{barticle}
\endbibitem

\bibitem{Pascale2012}
\begin{barticle}
\bauthor{\bsnm{Pascale}, \binits{E.}},
\bauthor{\bparticle{et} \bsnm{al.}}:
\batitle{The balloon-borne large-aperture submillimeter telescope for polarimetry-blastpol: performance and results from the 2010 antarctic flight}.
\bjtitle{Ground-based and Airborne Telescopes IV. Proceedings of the SPIE}
\bvolume{8444},
\bfpage{844415}
(\byear{2012}).
\doiurl{10.1117/12.927211}
\end{barticle}
\endbibitem

\bibitem{Bernard2016}
\begin{barticle}
\bauthor{\bsnm{Bernard}, \binits{J.P.}},
\bauthor{\bparticle{et} \bsnm{al.}}:
\batitle{Pilot: a balloon-borne experiment to measure the polarized fir emission of dust grains in the interstellar medium}.
\bjtitle{Experimental Astronomy}
\bvolume{42},
\bfpage{199}--\blpage{227}
(\byear{2016})
\end{barticle}
\endbibitem

\bibitem{Savini2012}
\begin{barticle}
\bauthor{\bsnm{Savini}, \binits{G.}},
\bauthor{\bsnm{Ade}, \binits{P.A.R.}},
\bauthor{\bsnm{Zhang}, \binits{J.}}:
\batitle{A new artificial material approach for flat thz frequency lenses}.
\bjtitle{Optics Express}
\bvolume{20}(\bissue{23}),
\bfpage{25766}--\blpage{25773}
(\byear{2012})
\end{barticle}
\endbibitem

\bibitem{Pisano2013}
\begin{barticle}
\bauthor{\bsnm{Pisano}, \binits{G.}},
\bauthor{\bsnm{Ng}, \binits{M.W.}},
\bauthor{\bsnm{Ozturk}, \binits{F.}},
\bauthor{\bsnm{Maffei}, \binits{B.}},
\bauthor{\bsnm{Haynes}, \binits{V.}}:
\batitle{Dielectrically embedded flat mesh lens for millimeter waves applications}.
\bjtitle{Applied Optics}
\bvolume{52},
\bfpage{2218}
(\byear{2013}).
\doiurl{10.1364/AO.52.002218}
\end{barticle}
\endbibitem

\bibitem{Hargrave2014}
\begin{barticle}
\bauthor{\bsnm{Hargrave}, \binits{P.}},
\bauthor{\bparticle{et} \bsnm{al.}}:
\batitle{Refractive telescope systems for future cosmic microwave background polarimetry experiments}.
\bjtitle{Ground-based and Airborne Telescopes V. Proceedings of the SPIE}
\bvolume{9153},
\bfpage{915314}
(\byear{2014}).
\doiurl{10.1117/12.2054634}
\end{barticle}
\endbibitem

\bibitem{Fox2017}
\begin{barticle}
\bauthor{\bsnm{Fox}, \binits{S.}}, \betal:
\batitle{Ismar: an airborne submillimetre radiometer}.
\bjtitle{Atmos. Meas. Tech}
\bvolume{10},
\bfpage{477}--\blpage{490}
(\byear{2017}).
\doiurl{10.5194/amt-10-477-2017}
\end{barticle}
\endbibitem

\bibitem{Bergada2016}
\begin{botherref}
\oauthor{\bsnm{Bergad\'{a}}, \binits{M.}}, et al.:
The ice cloud imager (ici) preliminary design and performance.
2016 14th Specialist Meeting on Microwave Radiometry and Remote Sensing of the Environment (MicroRad),
27--31
(2016).
\doiurl{10.1109/MICRORAD.2016.7530498}
\end{botherref}
\endbibitem

\bibitem{Zemax}
\begin{botherref}
ANSYS Zemax Opticstudio (Legacy).
\url{https://www.ansys.com/en-gb/products/optics/ansys-zemax-opticstudio}
\end{botherref}
\endbibitem

\bibitem{Marcuvitz}
\begin{botherref}
\oauthor{\bsnm{Marcuvitz}, \binits{N.}}:
Waveguide handbook
(1951)
\end{botherref}
\endbibitem

\bibitem{Ulrich67}
\begin{barticle}
\bauthor{\bsnm{Ulrich}, \binits{R.}}:
\batitle{Far-infrared properties of metallic mesh and its complementary structure}.
\bjtitle{Infrared Physics}
\bvolume{7},
\bfpage{37}--\blpage{55}
(\byear{1967}).
\doiurl{10.1016/0020-0891(67)90028-0}
\end{barticle}
\endbibitem

\bibitem{Rawcliffe67}
\begin{barticle}
\bauthor{\bsnm{Rawcliffe}, \binits{R.D.}},
\bauthor{\bsnm{Randall}, \binits{C.M.}}:
\batitle{Metal mesh interference filters for the far infrared}.
\bjtitle{Applied Optics}
\bvolume{6},
\bfpage{1353}--\blpage{1358}
(\byear{1967})
\end{barticle}
\endbibitem

\bibitem{Pisano2008}
\begin{barticle}
\bauthor{\bsnm{Pisano}, \binits{G.}},
\bauthor{\bsnm{Savini}, \binits{G.}},
\bauthor{\bsnm{Ade}, \binits{P.A.R.}},
\bauthor{\bsnm{Haynes}, \binits{V.}}:
\batitle{Metal-mesh achromatic half-wave plate for use at submillimeter wavelengths}.
\bjtitle{Applied Optics}
\bvolume{47},
\bfpage{6251}--\blpage{6256}
(\byear{2008}).
\doiurl{10.1364/AO.47.006251}
\end{barticle}
\endbibitem

\bibitem{Zhang2011}
\begin{barticle}
\bauthor{\bsnm{Zhang}, \binits{J.}},
\bauthor{\bparticle{et} \bsnm{al.}}:
\batitle{Polypropylene embedded metal mesh broadband achroatic half-wave plate for millimeter wavelengths}.
\bjtitle{Applied Optics}
\bvolume{50}(\bissue{21}),
\bfpage{3750}--\blpage{3757}
(\byear{2011})
\end{barticle}
\endbibitem

\bibitem{Moseley2017}
\begin{barticle}
\bauthor{\bsnm{Moseley}, \binits{P.}},
\bauthor{\bsnm{Savini}, \binits{G.}},
\bauthor{\bsnm{Zhang}, \binits{J.}},
\bauthor{\bsnm{Ade}, \binits{P.}}:
\batitle{Dual focus polarisation splitting lens}.
\bjtitle{Optics Express}
\bvolume{25}(\bissue{21}),
\bfpage{25363}--\blpage{25373}
(\byear{2017}).
\doiurl{10.1364/OE.25.025363}
\end{barticle}
\endbibitem

\bibitem{Pisano2023}
\begin{barticle}
\bauthor{\bsnm{Pisano}, \binits{G.}},
\bauthor{\bsnm{Dunscombe}, \binits{C.}},
\bauthor{\bsnm{Hargrave}, \binits{P.}},
\bauthor{\bsnm{Shitvov}, \binits{A.}},
\bauthor{\bsnm{Tucker}, \binits{C.}}:
\batitle{Thin flexible multi-octave metamaterial absorber for millimeter wavelengths}.
\bjtitle{Applied Optics}
\bvolume{62},
\bfpage{2317}
(\byear{2023}).
\doiurl{10.1364/AO.478842}
\end{barticle}
\endbibitem

\bibitem{Pisano2018}
\begin{barticle}
\bauthor{\bsnm{Pisano}, \binits{G.}},
\bauthor{\bsnm{Shitvov}, \binits{A.}},
\bauthor{\bsnm{Moseley}, \binits{P.}},
\bauthor{\bsnm{Tucker}, \binits{C.}},
\bauthor{\bsnm{Savini}, \binits{G.}},
\bauthor{\bsnm{Ade}, \binits{P.A.R.}}:
\batitle{Development of large-diameter flat mesh-lenses for millimetre wave instrumentation}.
\bjtitle{Ground-based and Airborne Telescopes V. Proceedings of the SPIE}
\bvolume{10708},
\bfpage{107080}
(\byear{2018}).
\doiurl{10.1117/12.2314063}
\end{barticle}
\endbibitem

\bibitem{Braithwaite}
\begin{barticle}
\bauthor{\bsnm{Braithwaite}, \binits{C.}},
\bauthor{\bparticle{et} \bsnm{al.}}:
\batitle{A study on the performance of half-wave plates designed for sub-mm studies}.
\bjtitle{Ground-based and Airborne Telescopes V. Proceedings of the SPIE}
(\byear{2024}).
\doiurl{10.1117/12.3019262}
\end{barticle}
\endbibitem

\bibitem{Moseley2018}
\begin{barticle}
\bauthor{\bsnm{Moseley}, \binits{P.}},
\bauthor{\bsnm{Savini}, \binits{G.}},
\bauthor{\bsnm{Ade}, \binits{P.}}:
\batitle{Large aperture metal-mesh lenses for thz astronomy}.
\bjtitle{Proc. of the 12th European Conference on Antennas and Propagation (EuCAP 2018)}
(\byear{2018}).
\doiurl{10.1049/cp.2018.0604}
\end{barticle}
\endbibitem

\bibitem{TKtiles}
\begin{botherref}
\oauthor{\bsnm{Jussi~Säily}, \binits{A.V.R.}}:
Studies on Specular and Non-specular Reflectivities of Radar Absorbing Materials (RAM) at Submillimetre Wavelengths
(February 2003).
\url{https://www.terahertz.co.uk/images/tki/RAM/HUT_RAM.pdf}
\end{botherref}
\endbibitem

\end{thebibliography}

\end{document}